\documentclass[9pt,conference]{IEEEtran}
\usepackage[T1]{fontenc}

\usepackage[preprint]{waspaa25} 

\usepackage{bm} 
\usepackage{booktabs}
\usepackage{siunitx}
\usepackage{color}
\usepackage{subcaption}
\usepackage{tablefootnote}

\title{Room-acoustic simulations as an alternative to measurements for audio-algorithm evaluation}

\name{Georg Götz$^{1*}$,
      Daniel Gert Nielsen$^{1}$,
      Steinar Gu\dh jónsson$^{1}$,
      Finnur Pind$^{1}$\thanks{The authors would like to thank Haukur Hafsteinsson, Hermes Sampedro Llopis, and Philip Puzalowski for their support while setting up the datasets.}}
\address{$^{1}$Treble Technologies, Reykjavík, Iceland\\ $^{*}$Correspondence: georg.goetz@treble.tech}

\begin{document}

\maketitle

\begin{abstract}
Audio-signal-processing and audio-machine-learning (ASP/AML) algorithms are ubiquitous in modern technology like smart devices, wearables, and entertainment systems. Development of such algorithms and models typically involves a formal evaluation to demonstrate their effectiveness and progress beyond the state-of-the-art. Ideally, a thorough evaluation should cover many diverse application scenarios and room-acoustic conditions. However, in practice, evaluation datasets are often limited in size and diversity because they rely on costly and time-consuming measurements. This paper explores how room-acoustic simulations can be used for evaluating ASP/AML algorithms. To this end, we evaluate three ASP/AML algorithms with room-acoustic measurements and data from different simulation engines, and assess the match between the evaluation results obtained from measurements and simulations. The presented investigation compares a numerical wave-based solver with two geometrical acoustics simulators. While numerical wave-based simulations yielded similar evaluation results as measurements for all three evaluated ASP/AML algorithms, geometrical acoustic simulations could not replicate the measured evaluation results as reliably.

\end{abstract}

\section{Introduction}
\label{sec:intro}
Modern technology has become deeply embedded in our daily lives through smart devices, wearables, and entertainment systems like virtual reality, advanced gaming platforms, and home theatre. These systems rely heavily on audio-signal-processing and audio-machine-learning (ASP/AML) algorithms to deliver seamless and immersive user experiences. The development of such algorithms, coupled with rigorous evaluation and benchmarking against the state-of-the-art, is essential for creating successful and competitive products. As these technologies become more sophisticated and context-aware, assessing algorithm performance in realistic acoustic environments while accounting for device specifics becomes critical to ensure their robustness and generalizability.

However, many typical evaluation datasets are limited in scale and diversity because they rely on costly, time-consuming, and labour-intensive room-acoustic measurements. Although recent advances in robotics have introduced promising possibilities for automating parts of this process using robotic fixtures \cite{Cmejla2021MirageDatabase,Koyama2021MeshRIR} or mobile platforms \cite{Goetz2021ARTSRAM,Stolz2023AutonomousRoboticPlatformSRIR,Goetz2023AutonomousRoomAcousticMeasurementsRRTGaussianProc}, scaling measurements to many rooms with varied geometries, materials, and furnishings remains a significant logistical and practical challenge. Consequently, the demand is increasing for scalable alternatives to room-acoustic measurements that enable ASP/AML algorithm evaluation across a more diverse range of acoustic environments.

Room-acoustic simulations are a scalable tool to generate large amounts of synthetic data under controlled, reproducible, and diverse acoustic conditions, making them a promising alternative to measurements when evaluating ASP/AML algorithm performance. A variety of simulation techniques exist \cite{Vorlaender2013ComputerSimulationsRoomAcoustics}, ranging from simple geometrical acoustics approaches like the image-source method or ray tracing~\cite{Savioja2015OverviewGA} to more advanced wave-based approaches \cite{Hamilton2016FiniteDifferenceVolumeWaveBasedRoomAcousticsThesis,Pind2019SpectralElementRoomAcousticSimulation,Prinn2023ReviewFEMRoomAcoustics}. Several studies have validated room-acoustic simulations under controlled conditions, and assessed their numerical and perceptual similarity to measurements~\cite{Brinkmann2019RoundRobinOnRoomAcousticalSimulationAndAuralization,Cosnefroy2023PhysicallyAccurateBinauralReproductionFromWaveBasedSimulation,Gudjonsson2025ValidationBroadbandWavebased,Llopis2025ValidationStudyFurnished,Puzalowski2024OptimizingSmallRoomWithTreble}. However, it remains uncertain whether the currently achieved numerical and perceptual fidelity is sufficient to replace room-acoustic measurements in practical ASP/AML algorithm evaluation. Moreover, the required simulation accuracy for this purpose is not well understood yet.

This paper investigates whether room-acoustic simulations can reliably replace physical measurements when evaluating ASP/AML algorithms. We assess three algorithms using measured data and simulations from one wave-based and two geometrical acoustics solvers. By comparing several performance metrics of the investigated ASP/AML algorithms across the different evaluation datasets, we examine how closely simulation-based results align with those from measurements.

The remainder of this paper is structured as follows. Section \ref{sec:background} summarizes different room-acoustic simulation concepts and techniques that are relevant for this study. Section \ref{sec:experiment_setup} describes the utilized datasets and how they are used to evaluate the investigated ASP/AML algorithms. Section \ref{sec:results} presents the results. Section \ref{sec:outlook} provides an outlook on how highly scalable simulations can provide more insights into algorithm performance than limited measurement datasets. Section~\ref{sec:conclusions} concludes the paper.



\section{Background}
\label{sec:background}
The sound field in rooms can be simulated with computers in various ways, which can be grouped according to two major room-acoustic simulation paradigms. 

The first paradigm is called ``Geometrical Acoustics (GA)'' and assumes that sound propagates through the environment as rays, thus neglecting its wave properties \cite{Savioja2015OverviewGA}. The GA paradigm includes widely used approaches like ray tracing, ray radiosity (``diffuse rain''), and the image-source method \cite{Vorlaender2008Auralization, Heinz1993BinauralRoomSimulationWithAdditionScattering, Schroeder2007FastReverberationEstimator}. However, the ray assumption makes it challenging to model wave effects like diffraction and interference at lower frequencies, for which the wavelength is large compared to the surfaces in the simulated model \cite{Torres2001ComputationEdgeDiffractionRoomAcousticSimulations,Schissler2014HighOrderDiffractionAndDiffuseReflectionsForInteractiveSoundPropagation, Brinkmann2019RoundRobinOnRoomAcousticalSimulationAndAuralization}.

The second paradigm comprises simulation techniques that numerically solve the wave equation, thus inherently modelling all wave phenomena. Such simulation techniques are commonly classified as ``wave-based simulations'' in the room-acoustics community. Different wave-based methods have been applied in room acoustics \cite{BottelDooren1995FDTDSimulationRoomAcoustic,Hamilton2016FiniteDifferenceVolumeWaveBasedRoomAcousticsThesis,Pind2019SpectralElementRoomAcousticSimulation,Prinn2023ReviewFEMRoomAcoustics}, with the discontinuous Galerkin finite-element method (DG-FEM)~\cite{Wang2019DiscontinuousGalerkinRoomAcoustics,Pind2020TimeDomainSimExtendedReactingPorousDGFEM} attracting significant attention recently due to its high scalability that allows massively parallel simulations on GPU clusters~\cite{Melander2024MassivelyParallelGalerkinRoomAcoustics}. These advances enable the simulation of broader frequency ranges, reducing the high computational costs that previously limited broadband wave-based simulations \cite{Siltanen2010RaysOrWavesStrenghtsAndWeaknesses}.

\begin{table*}
\caption{Overview of different datasets investigated in this study. Measurements were conducted in all rooms, and the simulation approaches outlined in Table \ref{table:overview_sims} were used to replicate them.}
\label{table:overview_datasets}
\centering
\begin{tabular}{lllll}
\hline
\textbf{Dataset} &
  \textbf{Room} &
  \textbf{Description} &
  \textbf{Source-receiver configurations} \\ \midrule
Bricks &
  \begin{tabular}[t]{@{}l@{}}Room 1: Lab room, \SI{80}{\meter^3} \\ (\SI{7}{\meter} $\times$ \SI{4.5}{\meter} $\times$ \SI{2.5}{\meter})\end{tabular} &
  \begin{tabular}[t]{@{}l@{}}Including piles of \SI{40}{\centi\meter} $\times$ \SI{40}{\centi \meter} bricks at multiple\\ locations and \SI{100}{\milli\meter} thick stonewool absorbers on\\ the walls.\vspace{0.25em} \\ Average reverberation time: \SI{0.6}{\second}\end{tabular} &
  \begin{tabular}[t]{@{}l@{}}20 RIRs (10 receivers $\times$ 2 sources) \\ Receivers: GRAS 1/2'' free-field microphones\\ Sources: Avantone MixCubes Active\end{tabular}  \\ \hline
Furniture &
  \begin{tabular}[t]{@{}l@{}}Room 1: Lab room, \SI{80}{\meter^3} \\ (\SI{7}{\meter} $\times$ \SI{4.5}{\meter} $\times$ \SI{2.5}{\meter})\end{tabular} &
  \begin{tabular}[t]{@{}l@{}}Including typical living room furniture and 100 mm \\  thick stonewool absorbers on the walls. \vspace{0.25em} \\ Average reverberation time \SI{0.5}{\second} \\ \end{tabular} &
  \begin{tabular}[t]{@{}l@{}}6 RIRs (3 receivers $\times$ 2 sources)\\  Receivers: GRAS 1/2'' free-field microphones\\ Sources: Avantone MixCubes Active\end{tabular}  \\ \hline
\begin{tabular}[t]{@{}l@{}}Variable\\ absorption\end{tabular} &
  \begin{tabular}[t]{@{}l@{}}Room 1: Lab room, \SI{80}{\meter^3} \\ (\SI{7}{\meter} $\times$ \SI{4.5}{\meter} $\times$ \SI{2.5}{\meter})\end{tabular} &
  \begin{tabular}[t]{@{}l@{}}Including different stonewool absorber configurations \\ on the walls, resulting in three acoustic conditions. \vspace{0.25em} \\ Average reverberation time, condition 1: \SI{0.3}{\second} \\ Average reverberation time, condition 2: \SI{0.6}{\second} \\ Average reverberation time, condition 3: \SI{0.9}{\second} \end{tabular} &
  \begin{tabular}[t]{@{}l@{}}12 RIRs (2 receivers $\times$ 2 sources $\times$ 3 conditions)\\ Receivers: GRAS 1/2'' free-field microphones\\ Sources: Avantone MixCubes Active\end{tabular} \\ \hline
Studio &
  \begin{tabular}[t]{@{}l@{}}Room 2: Studio, \SI{38}{\meter^3}\\ (\SI{4.8}{\meter} $\times$ \SI{3.2}{\meter} $\times$ \SI{2.5}{\meter})\end{tabular} &
  \begin{tabular}[t]{@{}l@{}}Empty studio room in a historic old building.\\ Solid exterior walls, lightweight construction\\ interior walls, laminate floor. \vspace{0.25em} \\ Average reverberation time: \SI{1.43}{\second} \end{tabular} &
  \begin{tabular}[t]{@{}l@{}}34 RIRs (17 receivers $\times$ 2 sources)\\ Receivers: NTI MA220 Class 1 microphones\\ Sources: ATC SCM25 A MK2\end{tabular} \\ \hline
\end{tabular}
\end{table*}
\begin{figure*}
  \begin{subfigure}[b]{0.33\textwidth}
    \includegraphics[width=\linewidth]{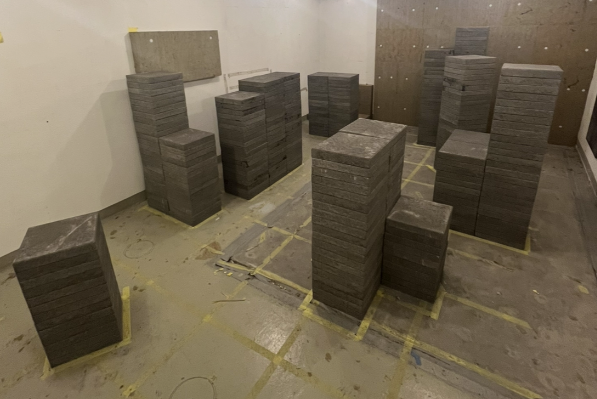}
    \caption{Lab room with bricks.}
    \label{subfig:measured_room}
  \end{subfigure}
  \begin{subfigure}[b]{0.33\textwidth}
    \includegraphics[width=\linewidth]{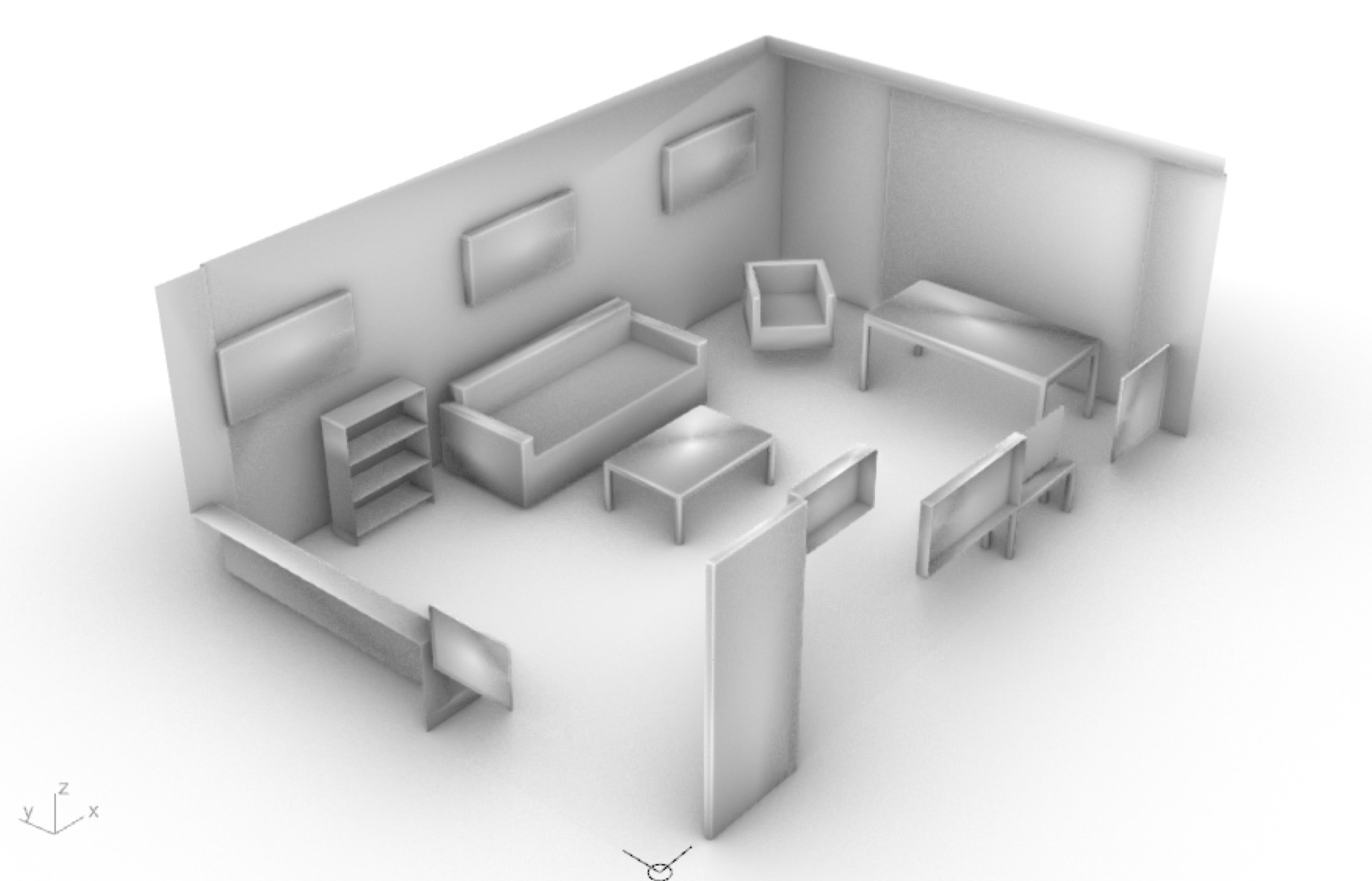}
    \caption{Lab room with furniture.}
    \label{subfig:furnished_room}
  \end{subfigure}
  \begin{subfigure}[b]{0.33\textwidth}
    \includegraphics[width=\linewidth]{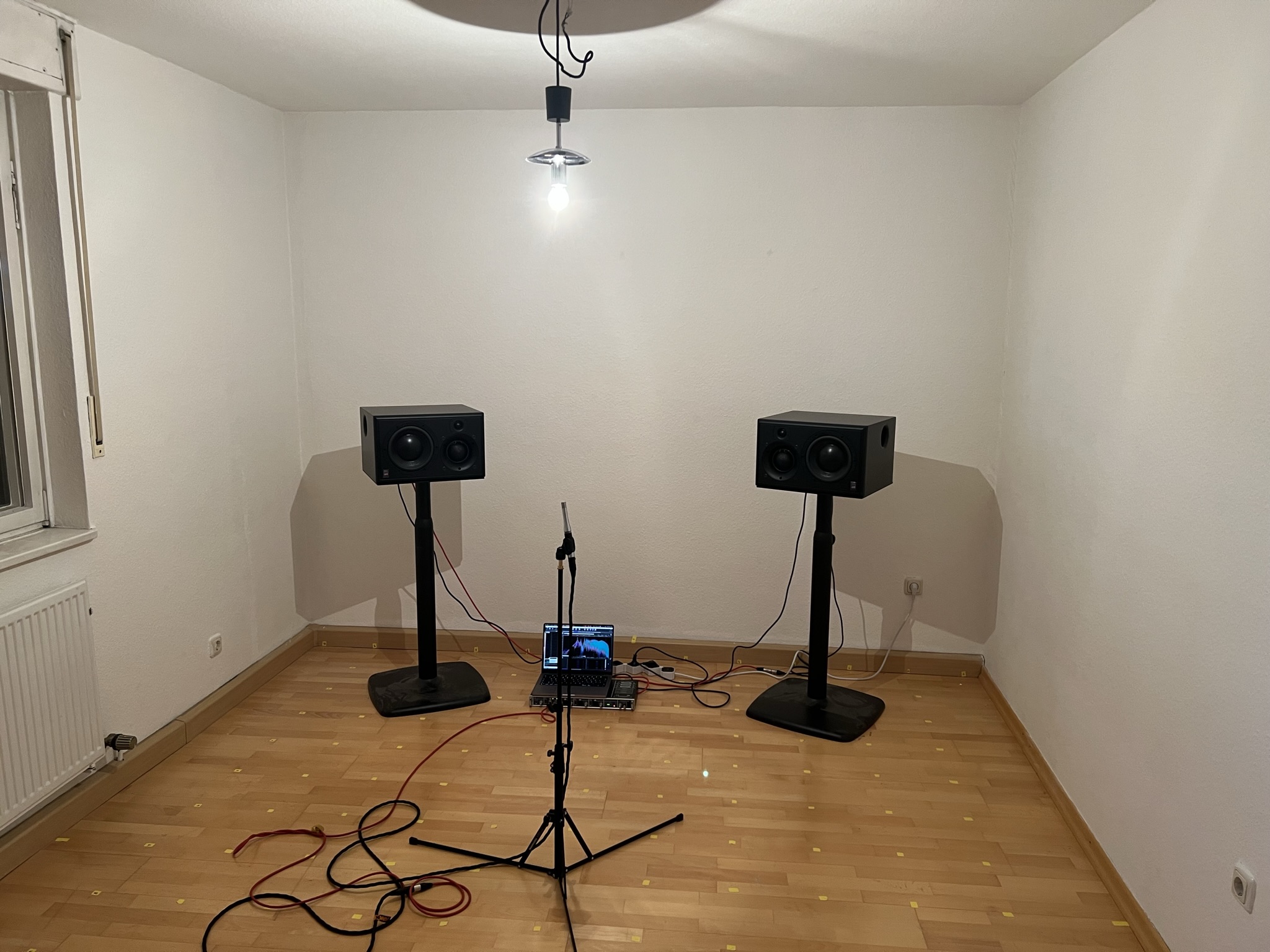}
    \caption{Studio.}
    \label{subfig:studio}
  \end{subfigure}
  


  
  \caption{Different room configurations (see Table \ref{table:overview_datasets}) were physically measured and then replicated with simulations.} 
  \label{fig:setup}
\end{figure*}

\section{Experiment setup}
\label{sec:experiment_setup}
The experiment presented in this paper investigates whether simulations can replace room-acoustic measurements when evaluating ASP/AML algorithms, and whether accurate modelling of wave effects plays a crucial role in that context. To this end, we evaluate three ASP/AML algorithms with data from measurements and three different simulations. The following section provides further details on the experiment.

\subsection{Datasets}
\label{subsec:datasets}

Table \ref{table:overview_datasets} summarizes the four room-acoustic measurement datasets that served as the baseline for this study. The measurements took place in two rooms with four different configurations, amounting to 72 RIRs in total. Figure \ref{fig:setup} illustrates three of the four room configurations. Measurements for the fourth dataset were conducted in the empty lab room with different configurations of stonewool absorbers \mbox{on the walls.} 

Three different simulation approaches were used to replicate the measurements. As summarized in Table \ref{table:overview_sims}, our study featured a wave-based simulation and GA simulations from two different simulators. All simulations used the same 3D model of the measured rooms including all geometrical details like bricks or furniture. The surface impedance of the absorbers in the lab room was determined with an impedance tube according to ISO 10534-2 \cite{ISO10534ImpedanceTube}, while the absorption characteristics of all other surfaces were obtained from established material databases. While the DG-FEM and GA-RR simulators support complex surface impedances, the GA-RT relied on energy absorption coefficients. In the DG-FEM simulation, the simulated frequency range was capped at \SI{7}{\kilo\Hz} to limit the computational cost. Consequently, the GA simulations and measurements were low-pass filtered at \SI{7}{\kilo\Hz} to make the RIRs comparable. All simulations and measurements were resampled to a sampling frequency of \SI{16}{\kilo\Hz}.


In addition to different simulation paradigms, the simulations also modelled the loudspeaker source directivity with different levels of accuracy. While DG-FEM simulations with the Treble SDK can apply a normal velocity directly to the membrane layer of the loudspeaker in the simulation model, GA simulations rely on predefined directivity patterns. Therefore, in the \mbox{GA-RR} simulation, we sample the boundary-velocity-source directivity pattern in \ang{5} steps, whereas the \mbox{GA-RT} simulation relies on an approximate loudspeaker directivity with a cardioid pattern. These two approaches reflect the typically used source modelling methods in the respective simulation engines. 

The numerical match between measurements and DG-FEM simulations from the lab room was evaluated in previous studies \cite{Gudjonsson2025ValidationBroadbandWavebased,Llopis2025ValidationStudyFurnished}. A similar study exists for the studio room \cite{Puzalowski2024OptimizingSmallRoomWithTreble}. Additional comparisons between measurements and the other simulations can be found on the companion page of this paper\footnote{The link will be inserted here upon acceptance of the paper.}.

\begin{table}[t]
\centering
\caption{Overview of different simulation approaches that were used to replicate the measurement dataset.}
\label{table:overview_sims}
\begin{tabular}{@{}lp{0.8\columnwidth}@{}}
\toprule
\textbf{Name} & \textbf{Simulation details}                                          \\ \midrule
DG-FEM       & \begin{tabular}[t]{@{}p{0.8\columnwidth}@{}}Wave-based: Discontinuous Galerkin finite-element method \cite{Wang2019DiscontinuousGalerkinRoomAcoustics,Pind2020TimeDomainSimExtendedReactingPorousDGFEM, Melander2024MassivelyParallelGalerkinRoomAcoustics}
                \begin{itemize}
                \item Implemented with the Treble SDK\tablefootnote{\label{footnote:treble}\url{https://www.treble.tech/software-development-kit}}
                \item Surfaces modelled with complex acoustic impedances
                \item Directional source modelling: boundary velocity source
                \item Upper simulation frequency: \SI{7}{\kilo\Hz}
                \end{itemize}\end{tabular} \vspace{-1em} \\ \hline
GA-RR        & \begin{tabular}[t]{@{}p{0.8\columnwidth}@{}} Geometrical Acoustics: Ray radiosity \cite{Heinz1993BinauralRoomSimulationWithAdditionScattering,Schroeder2007FastReverberationEstimator}
                \begin{itemize}
                \item Implemented with the Treble SDK$^\text{\ref{footnote:treble}}$
                \item Surfaces modelled with complex acoustic impedances
                \item Directional source modelling: fitted pattern from boundary velocity source
                \end{itemize}\end{tabular} \vspace{-1em} \\ \hline
GA-RT       & \begin{tabular}[t]{@{}p{0.8\columnwidth}@{}} Geometrical Acoustics: Ray tracing \cite{Vorlaender2008Auralization}
                \begin{itemize}
                \item Implemented with PyRoomAcoustics \cite{Scheibler2018PyRoomAcousticsICASSP}
                \item Surfaces modelled with energy absorption coefficients
                \item Directional source modelling: cardioid directivity pattern
                \end{itemize}\end{tabular} \vspace{-1em} \\ \bottomrule
\end{tabular}
\end{table}

\subsection{Evaluated algorithms}
We used the previously described datasets to evaluate three different ASP/AML algorithms. The first algorithm is a traditional signal processing algorithm for speech dereverberation based on the weighted prediction error (WPE) \cite{Yoshioka2012MultichannelLinearPredictionDereverb}. In our experiment, we used the offline implementation detailed in~\cite{Drude2018NaraWPEPythonPackageWeightedPredictionDereverb}. The WPE-based dereverberation algorithm requires multiple reverberant speech recordings from the same environment. Therefore, for each ``main RIR'' in each evaluation dataset, we randomly picked multiple ``support RIRs'' from the same dataset to support the dereverberation. The number of support RIRs varied between $4$ and $12$ depending on the investigated room configuration. This process is repeated for all RIRs of the dataset, such that every RIR is ``main RIR'' once. The dereverberation performance is then only evaluated for the ``main RIRs'', thus yielding as many data points as there are RIRs in the dataset. 

The second algorithm is a neural network for speaker-distance estimation (SDE) from reverberant speech \cite{Neri2024SpeakerDistanceEstimation}. Single-channel SDE is a challenging task that relies on accurate modelling of temporal and spectral signal characteristics. This property makes it an interesting task for assessing the fidelity of room-acoustic data generation, as showcased in the Generative Data Augmentation Challenge \cite{Lin2025GenDARAChallenge}. Our experiment used the official source code\footnote{\url{https://github.com/michaelneri/audio-distance-estimation}} for training and applying the network that was released together with the algorithm's publication. We trained the network on a custom reverberant speech dataset. To this end, we combine four datasets of measured RIRs to cover diverse room-acoustic conditions and source-receiver distances. The combined dataset includes the Arni \cite{Prawda2022CalibratingSabineAndEyring}, Motus \cite{Goetz2021MotusDatasetPaper}, MIT Survey~\cite{Traer2016StatisticsNaturalReverberationEnablePerceptualSeparationSoundSpace}, and C4DM dataset \cite{Stewart2010DatabaseOfOmniAndBformatRIRs}. The measured RIRs from the combined dataset are then convolved with anechoic speech signals from the VCTK dataset~\cite{VCTKDataset}, resulting in a training dataset that amounts to approximately \num{30000} reverberant speech signals in total. After training the SDE network, we stored the resulting network weights as a checkpoint that was subsequently evaluated with the different evaluation datasets outlined in Sec. \ref{subsec:datasets}. In the evaluation, RIRs from the evaluation datasets were convolved with speech from the TSP speech database \cite{Kabal2018TSPspeechdatabase}. The resulting reverberant speech signals were then processed with the pre-trained network. 

The third algorithm is a diffusion-based generative neural network for speech dereverberation \cite{Richter2023SpeechEnhancementDereverberationDiffusionSGMSE}. We used the official source code\footnote{\url{https://github.com/sp-uhh/sgmse}} that was released together with the publication for applying the network. Additionally, we obtained the pretrained WSJ0-REVERB checkpoint from the official repository, and used in our evaluation. The evaluation of the pre-trained model followed the same procedure as described for the second algorithm.

\subsection{Performance metrics of the evaluated algorithms}
\label{subsec:performance_metrics}
The performance of the dereverberation algorithms was evaluated using several standard objective metrics. We employed the Perceptual Evaluation of Speech Quality (PESQ) metric \cite{itu_pesq_2001} to assess the perceived quality of dereverberated speech. PESQ takes values ranging from 1 to 5, where higher scores indicate better quality. To measure speech intelligibility of the dereverberated speech, the Extended Short-Time Objective Intelligibility (ESTOI) metric~\cite{taal_stoi_2010,jensen_taal_estoi_2016} was utilized, providing scores between 0 (unintelligible) and~1~(perfectly intelligible). Additionally, the fidelity of dereverberation was also quantified using the Scale-Invariant Signal-to-Distortion Ratio (SI-SDR) \cite{leroux_sisdr_2019}, where higher values in decibels (\si{\decibel}) indicate superior dereverberation performance. Finally, the SDE algorithm performance was assessed using the distance estimation error, calculated as the absolute error between the estimated and true source distances in meters, with lower values indicating better performance.

\section{Results}
\label{sec:results}
\begin{figure*}[tbh!]
  \centering
  \begin{subfigure}{\textwidth}
      \centerline{\includegraphics[width=\textwidth, trim={0 0 0 4em}, clip]{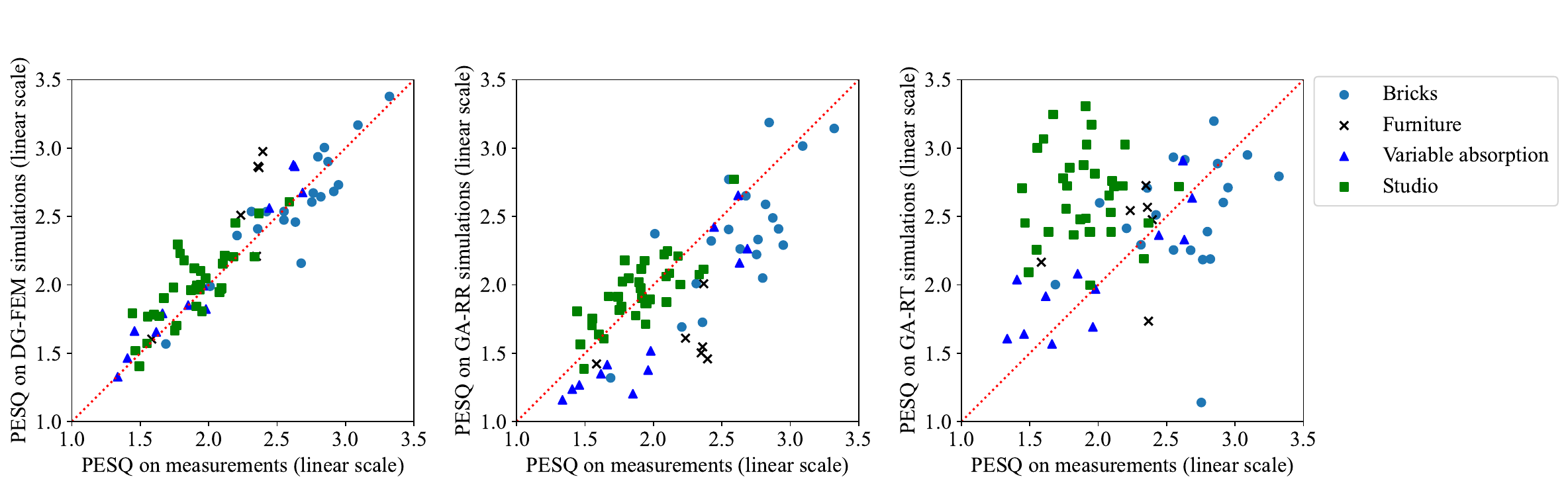}}
      \caption{WPE-based dereverberation algorithm}
      \label{subfig:scatter_dereverb_wpe}
  \end{subfigure}
  
\vspace{1em}

  \begin{subfigure}{\textwidth}
      \centerline{\includegraphics[width=\textwidth, trim={0 0 0 0em}, clip]{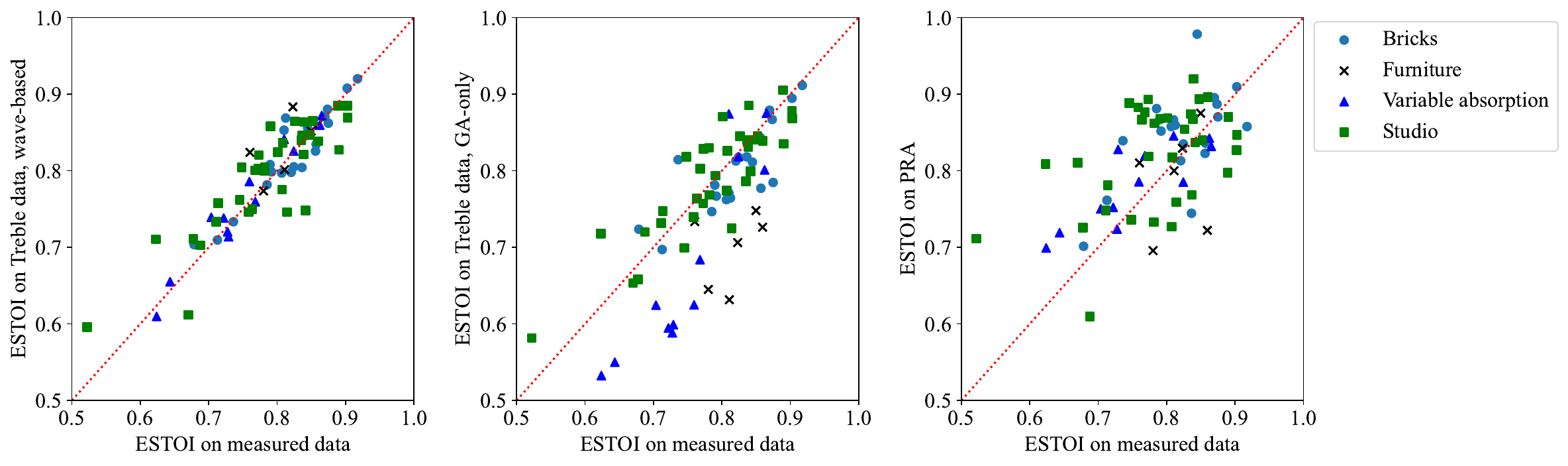}}
      \caption{Neural-network-based dereverberation algorithm}
      \label{subfig:scatter_dereverb_ml}
  \end{subfigure}
  
\vspace{1em}

  \begin{subfigure}{\textwidth}
    \centerline{\includegraphics[width=\textwidth, trim={0 0 0 4em}, clip]{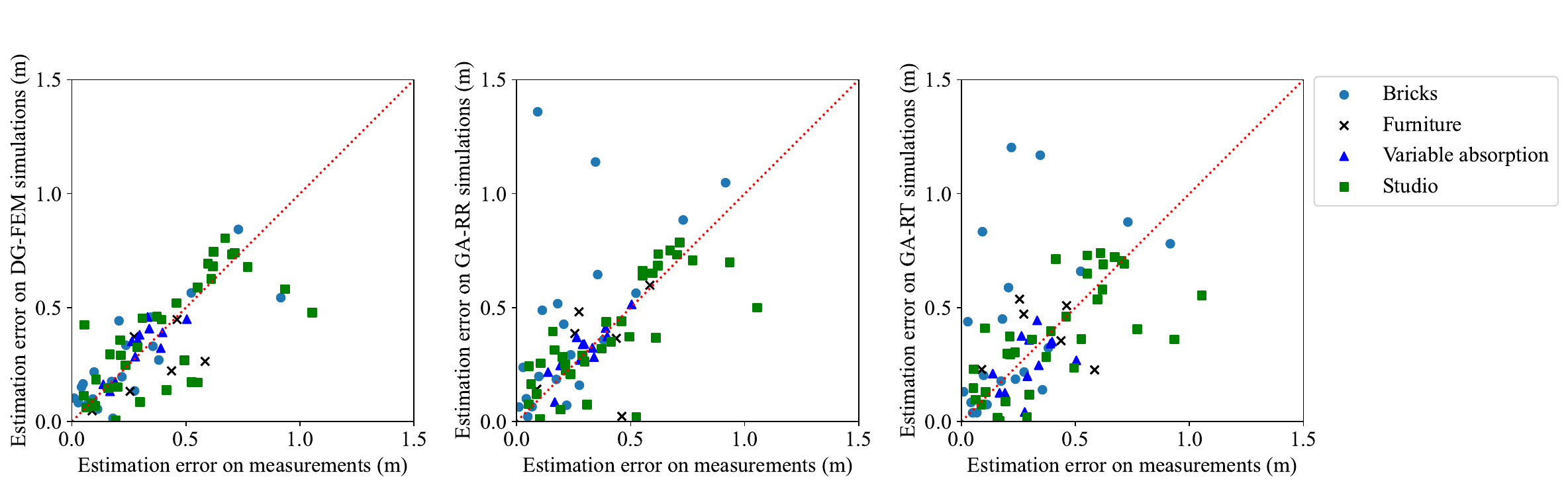}}
    \caption{Neural-network-based SDE algorithm}
  \label{subfig:scatter_sde_ml}
  \end{subfigure}
  \caption{Evaluation of ASP and AML algorithms with measured and simulated data. The scatter plots illustrate that DG-FEM simulations yield similar evaluation results as measurements, while GA simulations show larger deviations. The red dotted lines indicate where measured and simulated results would perfectly match.}
  \label{fig:scatter_eval}
\end{figure*}

Figure~\ref{fig:scatter_eval} compares the evaluation results obtained with measured and simulated data for the three investigated ASP/AML algorithms. The left plot in Fig.~\ref{subfig:scatter_dereverb_wpe} shows that PESQ scores obtained from the measured evaluation data almost coincide with the scores obtained from the DG-FEM simulations that replicated the measurements. In contrast, the evaluation results of both GA simulations deviate considerably from the evaluation results obtained from the measurements, as evidenced by the middle and right plot in Fig.~\ref{subfig:scatter_dereverb_wpe}. Similar trends can be observed for the evaluation of the neural-network-based dereverberation and SDE algorithms, as shown in Figs.~\ref{subfig:scatter_dereverb_ml} and \ref{subfig:scatter_sde_ml}, which depict ESTOI and the speech distance estimation errors obtained on the different evaluation datasets, respectively. 

In the following, we will use two metrics to quantify the match between the evaluation results obtained from measured and simulated data. The first metric is the Pearson correlation coefficient $\rho$ that measures the linear correlation between two datasets. Let $X = \{x_1, x_2, \ldots, x_N\}$ and $Y = \{y_1, y_2, \ldots, y_N\}$ denote the $N$ evaluation results for a certain performance metric (see Sec.~\ref{subsec:performance_metrics}) obtained from two datasets, respectively. With $\overline{x}$ and $\overline{y}$ being the means over the respective datasets, $\rho$ is calculated as
\begin{equation}
\rho = \frac{\sum\limits_{i=1}^{N} (x_i-\overline{x})(y_i-\overline{y})}{\sqrt{\sum\limits_{i=1}^{N}(x_i-\overline{x})^2\sum\limits_{i=1}^{N}(y_i-\overline{y})^2}} \,.
\end{equation}
It takes values between $-1$ and $1$, where the extremes correspond to exact negative and positive linear correlation, respectively, and a value of $0$ indicates no correlation. 

The second metric is the root mean squared error (RMSE) between the evaluation results obtained from two different datasets
\begin{equation}
    \mathrm{RMSE} = \sqrt{\frac{1}{N}\sum_{i=1}^{N} (x_i - y_i)^2} \,.
\end{equation}
Two data sets that yield exactly the same evaluation results would exhibit the lowest possible RMSE of 0.

\begin{table*}[t]
\centering
\caption{Match between the evaluation results obtained from measurements and different simulations. The table reports the Pearson correlation coefficient $\rho$ and the root mean squared error (RMSE) between measured evaluation results and the corresponding results obtained from DG-FEM, GA-RR, and GA-RT simulations. Results are compared for three different ASP/AML algorithms with their respective performance metrics. Across all metrics, DG-FEM simulations yield similar evaluation results as measurements, while geometrical acoustic simulations could not replicate the measured evaluation results as precisely.}
\label{tbl:eval_results}
\sisetup{
    reset-text-series = false, 
    text-series-to-math = true, 
    mode=text,
    tight-spacing=true,
    round-mode=places,
    round-precision=2,
    table-format=2.2,
    table-number-alignment=center,
    detect-weight=true,
    detect-family=true
}
\begin{tabular}{@{}l 
S[round-precision=2, table-format=1.2]
S[round-precision=2, table-format=1.2]
S[round-precision=2, table-format=1.2]
S[round-precision=2, table-format=1.2]
S[round-precision=2, table-format=1.2]
S[round-precision=2, table-format=1.2]
S[round-precision=2, table-format=1.2]
S[round-precision=2, table-format=1.2]
S[round-precision=2, table-format=1.2]
S[round-precision=2, table-format=1.2]
S[round-precision=2, table-format=1.2]
S[round-precision=2, table-format=1.2]
S[round-precision=2, table-format=1.2]
S[round-precision=2, table-format=1.2]
@{}}
\toprule
 &
  \multicolumn{2}{l}{\textbf{SDE (ML)}} &
  \multicolumn{6}{l}{\textbf{Dereverberation (ML)}} &
  \multicolumn{6}{l}{\textbf{Dereverberation (DSP)}} \\ \cmidrule(lr){2-3} \cmidrule(lr){4-9} \cmidrule{10-15} 
 &
  \multicolumn{2}{l}{Distance est. error} &
  \multicolumn{2}{l}{PESQ} &
  \multicolumn{2}{l}{ESTOI} &
  \multicolumn{2}{l}{SI-SDR} &
  \multicolumn{2}{l}{PESQ} &
  \multicolumn{2}{l}{ESTOI} &
  \multicolumn{2}{l}{SI-SDR} \\ \cmidrule(lr){2-3} \cmidrule(lr){4-5} \cmidrule(lr){6-7} \cmidrule(lr){8-9} \cmidrule(lr){10-11} \cmidrule(lr){12-13} \cmidrule{14-15}
         &   $\rho \uparrow$ &
  \text{RMSE $\downarrow$}&
  \text{$\rho \uparrow$} &
  \text{RMSE $\downarrow$} &
  \text{$\rho \uparrow$} &
  \text{RMSE $\downarrow$} &
  \text{$\rho \uparrow$} &
  \text{RMSE $\downarrow$} &
  \text{$\rho \uparrow$} &
  \text{RMSE $\downarrow$} &
  \text{$\rho \uparrow$} &
  \text{RMSE $\downarrow$} &
  \text{$\rho \uparrow$} &
  \text{RMSE $\downarrow$}     \\ \midrule
DG-FEM &
  \bfseries 0.76 &
  \bfseries 0.158 &
  \bfseries 0.922 &
  \bfseries 0.218 &
  \bfseries 0.909 &
  \bfseries 0.054 &
  \bfseries 0.747 &
  \bfseries 3.090 &
  \bfseries 0.91 &
  \bfseries 0.205 &
  \bfseries 0.90 &
  \bfseries 0.033 &
  \bfseries 0.73 &
  \bfseries 2.352 \\
GA-RR & 0.587 & 0.237 & 0.681 & 0.510 & 0.704 & 0.105 & 0.61 & 3.675  & 0.77 & 0.35 & 0.76 & 0.064 & 0.57 & 2.857 \\
GA-RT    & 0.514 & 0.25 & 0.278 & 0.810 & 0.454 & 0.168 & 0.227  & 5.71 & 0.14 & 0.697 & 0.56 & 0.072 & 0.1 & 4.192 \\ \bottomrule
\end{tabular}
\vspace{1em}
\end{table*}

Table \ref{tbl:eval_results} summarizes the match between the evaluation results obtained from measurements and different simulations using the previously defined metrics. For all three evaluated ASP/AML algorithms, and all of their respective performance metrics, DG-FEM simulations yield evaluation results that exhibit high correlation with the evaluation results obtained from measurements. Additionally, their deviation from the measured evaluation results is only small, as evidenced by the RMSE. In contrast, GA simulations do not yield as highly correlated evaluation results, and also exhibit significantly larger deviations from measured evaluation results in terms of RMSE. Furhtermore, GA-RR simulations yield evaluation results that match slightly better with measured evaluation results than those from GA-RT simulations 

\section{Outlook: Extending ASP/AML evaluation through highly scalable simulations}
\label{sec:outlook}
The previous section demonstrated that DG-FEM simulations can be used as an alternative to measurements when evaluating ASP/AML algorithms. Their high scalability enables massively parallel simulations on GPU clusters~\cite{Melander2024MassivelyParallelGalerkinRoomAcoustics}, making them a powerful tool to set up large and diverse evaluation datasets for benchmarking audio algorithms. For instance, the simulation setup described in Section \ref{sec:experiment_setup} can be easily extended by adding dense receiver grids, without significantly increasing the computational complexity. Such a simulation setup facilitates algorithm evaluation with a level of detail that would not be feasible with measurements due to practical and cost constraints. Furthermore, DG-FEM simulations could be used to scale up the evaluation to many different rooms, without requiring physical access to them or carrying equipment around. 

To make DG-FEM simulations more accessible for the research community, we plan to publish an extended version of the dataset outlined in this study upon acceptance of this paper. This dataset includes the RIRs used for the present study, alongside additional RIRs from a dense receiver grid. Moreover, we will include DG-FEM simulations from additional rooms.

\section{Conclusions}
\label{sec:conclusions}
This study investigated how room-acoustic simulations can be used for evaluating audio-signal-processing and audio-machine-learning algorithms. We evaluated one traditional signal processing and two machine-learning-based algorithms with several performance metrics, and showed that high-accuracy wave-based simulations yield similar evaluation results as measurements. In contrast, geometrical acoustic simulations could not replicate the measured evaluation results as precisely. These results show that wave-based simulations can be a reliable alternative to measurements during the development cycle of audio algorithms. Furthermore, highly scalable simulations can be leveraged to evaluate algorithms in a wide variety of acoustic conditions, which would not be viable when relying on costly and time-consuming measurements.

\bibliographystyle{IEEEtran}
\bibliography{refs25}

\end{document}